\documentclass{article}

\usepackage{arxiv}

\usepackage[utf8]{inputenc} 
\usepackage[T1]{fontenc}    
\usepackage{hyperref}       
\usepackage{url}            
\usepackage{booktabs}       
\usepackage{amsfonts}       
\usepackage{nicefrac}       
\usepackage{microtype}      
\usepackage{lipsum}		
\usepackage{graphicx}
\usepackage{natbib}
\usepackage{doi}
\usepackage{amsmath}

\title{Achromatic photonic tricouplers for application in nulling interferometry}

\author{
\href{https://orcid.org/0000-0002-0989-9302}{\includegraphics[scale=0.06]{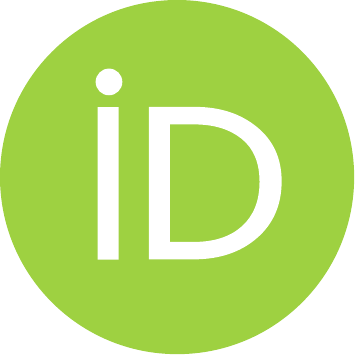}\hspace{1mm}Marc-Antoine Martinod}\thanks{email: \url{marc-antoine.martinod@sydney.edu.au}} \\
	Sydney Institute for Astronomy\\
	School of Physics\\
	The University of Sydney\\
	NSW 2006, Australia \\
	\And
	Peter Tuthill\\
	Sydney Institute for Astronomy\\
	School of Physics\\
	The University of Sydney\\
	NSW 2006, Australia \\
	\AND
	\href{https://orcid.org/0000-0001-5130-183X}{\includegraphics[scale=0.06]{orcid.pdf}\hspace{1mm}Simon Gross} \\
	MQ Photonics Research Centre\\
	Department of Physics and Astronomy\\
	Macquarie University\\
	Sydney, Australia \\
	\And
	\href{https://orcid.org/0000-0002-8352-7515}{\includegraphics[scale=0.06]{orcid.pdf}\hspace{1mm}Barnaby Norris}\\
	Sydney Institute for Astronomy\\
	School of Physics\\
	The University of Sydney\\
	NSW 2006, Australia \\
	\And
	David Sweeney \\
	Sydney Institute for Astronomy\\
	School of Physics\\
	The University of Sydney\\
	NSW 2006, Australia \\
	\And
	\href{https://orcid.org/0000-0002-6414-8739}{\includegraphics[scale=0.06]{orcid.pdf}\hspace{1mm}Michael J. Withford}\\
	MQ Photonics Research Centre\\
	Department of Physics and Astronomy\\
	Macquarie University\\
	Sydney, Australia \\	
}

\date{}

\renewcommand{\shorttitle}{\textit{Achromatic photonic tricouplers for application in nulling interferometry}}

\hypersetup{
pdftitle={Achromatic photonic tricouplers for application in nulling interferometry},
pdfsubject={photonics, astronomy, nulling},
pdfauthor={David S.~Hippocampus, Elias D.~Striatum},
pdfkeywords={First keyword, Second keyword, More},
}

\begin{document}
\maketitle
\copyright{2021 Optical Society of America. One print or electronic copy may be made for personal use only. Systematic reproduction and distribution, duplication of any material in this paper for a fee or for commercial purposes, or modifications of the content of this paper are prohibited.\\}

\begin{abstract}
Integrated-optic components are being increasingly used in astrophysics, mainly where accuracy and precision are paramount.
One such emerging technology is nulling interferometry that targets high contrast and high angular resolution.
Two of the most critical limitations encountered by nullers are rapid phase fluctuations in the incoming light causing instability in the interference and chromaticity of the directional couplers that prevent a deep broadband interferometric null.
We explore the use of a tricoupler designed by ultrafast laser inscription that solves both issues. 
Simulations of a tricoupler, incorporated into a nuller, result in order of a magnitude improvement in null depth.
\end{abstract}


\section{Introduction}
Integrated-optics and photonic technologies are becoming more widely used in astrophysics, particularly in high angular resolution and high contrast imaging.
Photonics-based solutions are smaller, lighter and cheaper than their bulk-optics equivalents, while at the same
time they can deliver functionality that is otherwise difficult, or even impossible to achieve (e.g. spatial filtering).
Once starlight is injected into photonic circuits, subsequent losses can be low and complex sequences of optical processing can be performed with no possibility for misalignment or drift. 
Thus, significant improvements in the instrument stability are possible.
Unprecedented precision has been reached by deploying such components in interferometry in the visible \cite{martinod2018, huby2012} and in the infrared domain \cite{gravity2017, 2019A&A...623L..11G}.
    
The detection and the characterization of exoplanets close to their host star, and particularly within the habitable zone, is one of the most pressing instrumental challenges faced by contemporary astronomy.
Such detections require a high angular resolution and ability to handle a high planet-to-star contrast ratio, the latter ranging from $10^{-4}$, for self-luminous hot exoplanets observed in the mid-infrared \cite{Marois2008}, to $10^{-10}$, for Earth-like exoplanets imaged in reflected light from their host star \cite{Schworer2015}.
Nulling interferometry fulfills both requirements by making an on-axis source (the star) destructively interfere while transmitting the light from an off-axis companion (a planet) \cite{Bracewell1978}.
The huge potential of integrated-optics to process the light for this observing technique has been demonstrated by the Guided-Light Interferometric Nulling Technology (GLINT) instrument \cite{lagadec2018, norris2020}.

GLINT is the first simultaneous multiple baseline nulling interferometer, also delivering photometric monitoring of the input beams and spectrally-dispersed fringe data.
It is deployed at the Subaru Telescope and integrated into the Subaru Coronagraphic Extreme Adaptive Optics system (SCExAO) \cite{Guyon2011, Jovanovic2013, Jovanovic2015} which provides wavefront correction for the atmospheric turbulence.
The present generation GLINT chip combines four beams arriving from four sub-apertures on the telescope pupil,
providing six non-redundant interferometric baselines spanning from 2.15 to 6.45~meters.
The light processing is performed within an integrated-optics single-mode component presently optimised to work over astronomical H-band. 
This chip has two main functions: firstly it coherently remaps the beams from the input 2-dimensional configuration into a more convenient output format (here, 1-dimensional), and secondly, the beams enter pairwise in directional couplers where interference occurs.
The directional couplers separate the nulled signal -- the on-axis starlight destructively interferes -- from the bright one -- the on-axis starlight constructively interferes, delivering these to two different output waveguides.
The ratio of these outputs give the \emph{null depth}, and during scientific use, this provides the observable from which astronomical measurements can be derived (for example, the contrast ratio of a companion object).
The photonic component also delivers photometric outputs to monitor the flux of each beam incident on the interferometric stage: a crucial signal for the accurate calibration of the outcome.
All outputs are spectrally dispersed with a resolving power of $\sim$\,160 before being  recorded on a sensitive infrared camera (C-Red\,2 from First Light Imaging).
The final data stream consists of a time-sequence of measured null depths that are processed according to the \emph{Numerical Self Calibration} (NSC) method \cite{hanot2011}.
This method employs a statistical analysis of the measured signals in order to recover the underlying null depth of the source.
This can ultimately be used to constrain the spatial brightness distribution of the target.

As presently configured, in-lab characterization of GLINT has shown that null depth, processed by NSC, reaches $10^{-3}$ with a precision of $10^{-4}$ \cite{martinod2021}.
While competitive within its niche of separation and contrast, these levels of performance are not sufficient as an operative exoplanetary imaging instrument reaching beyond the handful of most favourable target systems. 
The two main limitations to the performance are: (1) the residual fluctuations in the phase of the fringes due to imperfection in the correction provided by the adaptive optics (AO) system, and (2) the intrinsic chromaticity of the null depth due to the varying performance of the directional couplers across the observing band.
These issues could be addressed in a number of ways such as a dedicated fringe-tracking system or a supplementary light-processing stage in the photonic component \cite{Martin2014} to lock-in the dark fringe.
However, they require a dramatic up-scaling in size and complexity of the design and are unsuited for a pupil-remapping nuller like GLINT.
Here, a simple and compact solution, ideal for aperture-masking nulling interferometry and also compatible for long-baseline interferometry, is investigated.
It consists in a single integrated-optic device based on the tricoupler which solves both issues simultaneously.

The tricoupler is a photonic component which splits the pair of input beams into three interferometric outputs instead of two; previously mooted as a combiner for nulling and stellar interferometry \cite{weber2004, labeye2004} but never exploited.
It also has similarities with the Discrete Beam Combiners (DBC) \cite{minardi2010}, such as the capability to combine more than 2 apertures.
However, they have a significant difference: the DBC exhibits chromatic behaviour which can bias the measurement of interferometric observables \cite{saviauk2013} while the tricoupler can deliver a purely achromatic null depth.
Indeed, if the input beams are balanced and in anti-phase (orthodox injection for a nulling configuration) then simple geometrical symmetry requires that the nulled channel remains dark, regardless of any chromaticity of the coupling ratios of the device \cite{labeye2004}.
Furthermore, the device yields two other outputs configured in such a way that the differential flux between them yields a measurement of the phase delay at the input.
This immediately delivers the capability to perform simultaneous phase measurements along with the astrophysical observation (in the null channel) all in a single device.
The phase measurements will be sent to an active component (e.g. the existing segmented MEMS mirrors on GLINT \cite{norris2020, martinod2021}, a preceding adaptive optics system or active elements on the photonic chip itself \cite{martin2020}) to compensate for the piston due to the turbulence.
The configuration by which the tricoupler accomplishes fringe tracking is ideal: it removes any so-called {\it non-common path aberrations} which arise when the optical path to the fringe sensor differs from that to the science camera.
A 3-dimensional tricoupler for GLINT is now being manufactured by Ultrafast Laser Inscription method (ULI) \cite{Nolte2003, Gattass2008, Arriola2013, Gross2015} engineered as a simple drop-in replacement for the existing chip in the current optical layout \cite{martinod2021}.\\

This article reports the first numerical study on the improvement of the performance of a pupil-remapping nulling interferometer (GLINT) from the deployment of a tricoupler instead of a directional coupler.
The tricoupler's provision of simultaneous nulling and fringe-tracking operations can be exploited to yield improvement to the data quality, as revealed after processing based on the NSC method.
Section~\ref{sec:tricoupler} describes the characteristics of our tricoupler, explains the use of ULI technology and gives a comparison with the existing directional coupler.
Simulations comparing the performance of the tricoupler and the directional coupler as implemented in the GLINT nulling instrument are detailed in Section~\ref{sec:tricoupler_on_nulling}, with the results discussed in Section~\ref{sec:discussion}.

\section{Tricouplers, symmetry-enforced nulling and chromaticity}
\label{sec:tricoupler}
The tricoupler consists of three singlemode waveguides whose mode fields overlap over a defined interaction region.
For use in a nuller, the considered configuration is where only two of the input waveguides are illuminated as in a pair-wise interferometer.
The simplest geometries for the interaction region are those where three parallel guides are aligned side-by-side in a plane, or more interesting for GLINT, in a 3-D structure at the points of an equilateral triangle. 

For either of these cases, the tricoupler offers an intriguing and potentially valuable property to a nulling instrument.
If the incident electric field can be arranged to be strictly anti-symmetric -- injecting the left and right input channels with beams that are in anti-phase -- then the intrinsic symmetry of the structure means that none of the light is able to couple into the central guide.
For nulling, this center waveguide is exploited as the {\it null channel}, and because the inability of light to couple here is enforced by geometrical symmetry alone, the null so obtained should be both achromatic and robust against variations in the true coupling ratios obtained in a real fabricated device (so long, of course, as the symmetry is closely preserved). 
The ideal working tricoupler would have equal coupling ratios between all three guides which then have an output phase relationship of 120$^\circ$ to each other \cite{vance1994, xie2012}.
For the planar device to achieve this balance of the coupling ratios and phase delays across some operational working bandwidth, tapers have to be introduced to adiabatically alter the coupling along the light propagation direction \cite{vance1994}. 
This approach for instance was taken by Hsiao~\textit{et~al.} \cite{hsiao2010} to design a planar tricoupler for nulling interferometry, by Schneider~\textit{et~al.} \cite{schneider2000} for reducing the device's sensitivity to fabrication imperfections,
in the PIONIER instrument at VLTI for splitting the light in the beam combiner \cite{pionier2011, benisty2009}, and also by Lacour~\textit{et~al.} \cite{lacour2014} for combining nulling and closure phase.

The exploration beyond planar into 3~dimensional structures shows that further symmetries can be exploited as well as greater flexibility in tuning the coupling matrix, compared to the planar structure.
In particular, the equidistant arrangement of identical waveguides offered by an equilateral triangle naturally produces an equal splitting ratio between the two outer channels when used in the nulling configuration described above.
Early tricouplers, commonly referred to $3\times3$ couplers, were based on fused fibre technology and resulted in just such an equilateral triangle in the fused region. 
However, twisting during the fusion process can break the symmetry, altering the coupling ratio \cite{birks1992}.
In contrast, ULI enables the fabrication of 3D photonic components such as tricouplers with equilateral triangle coupling geometry \cite{Suzuki:06, Spagnolo2013, Chaboyer2015}. 
Unlike, fused-fibre tricouplers, they are free of twist and the geometry and waveguide parameters can be easily tailored to achieve the desired symmetric performance thus provide an achromatic-nulling and low-loss tricoupler.\\

\begin{figure}
    \centering
    \includegraphics[width=0.3\textwidth]{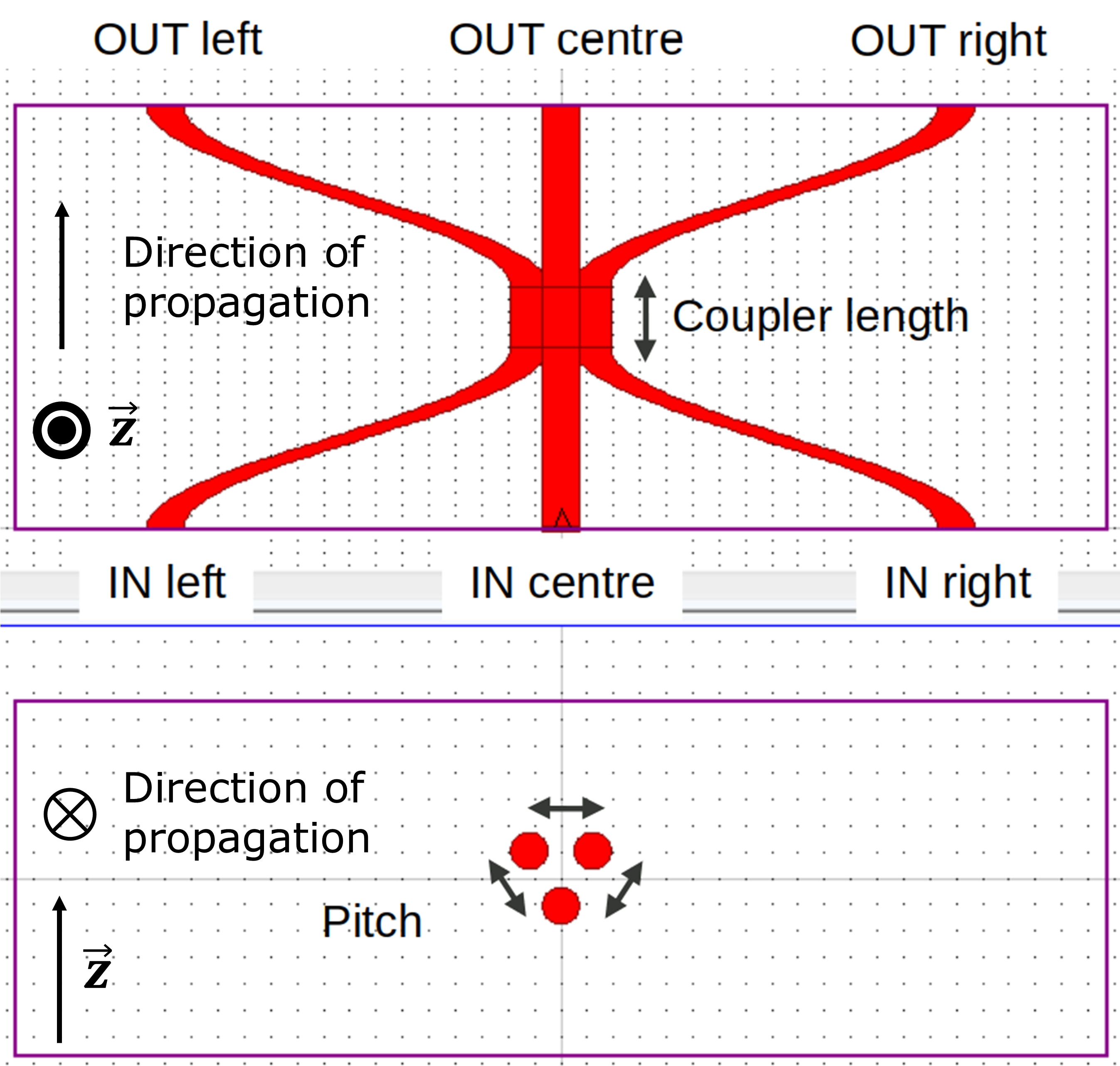}
    \caption{Schematic diagram of the simulated equilateral tricoupler with the labelled inputs and outputs (not on a uniform x,y scale). Top: top view of the tricoupler. Bottom: cross-section of the tricoupler.}
    \label{fig:tricoupler_scheme}
\end{figure}

A candidate idealised equilateral triangle tricoupler designed for GLINT was simulated using the  {\tt BeamProp} package within the {\tt RSoft} numerical simulation environment (Fig.~\ref{fig:tricoupler_scheme}). 
Assuming fused silica for the integrated-optics component (refractive index~$=1.444$), the coupling length of the waveguides is 1700~$\mu$m which provides equal splitting at 1.55~$\mu$m.
Beams are injected into the left and right inputs and the data consists of fluxes measured from the three outputs: the nulled signal is collected in the centre output, while the left and right outputs are used to recover the phase of the incoming beams to perform fringe tracking.\\

\begin{figure}
    \centering
    \includegraphics[width=0.48\textwidth]{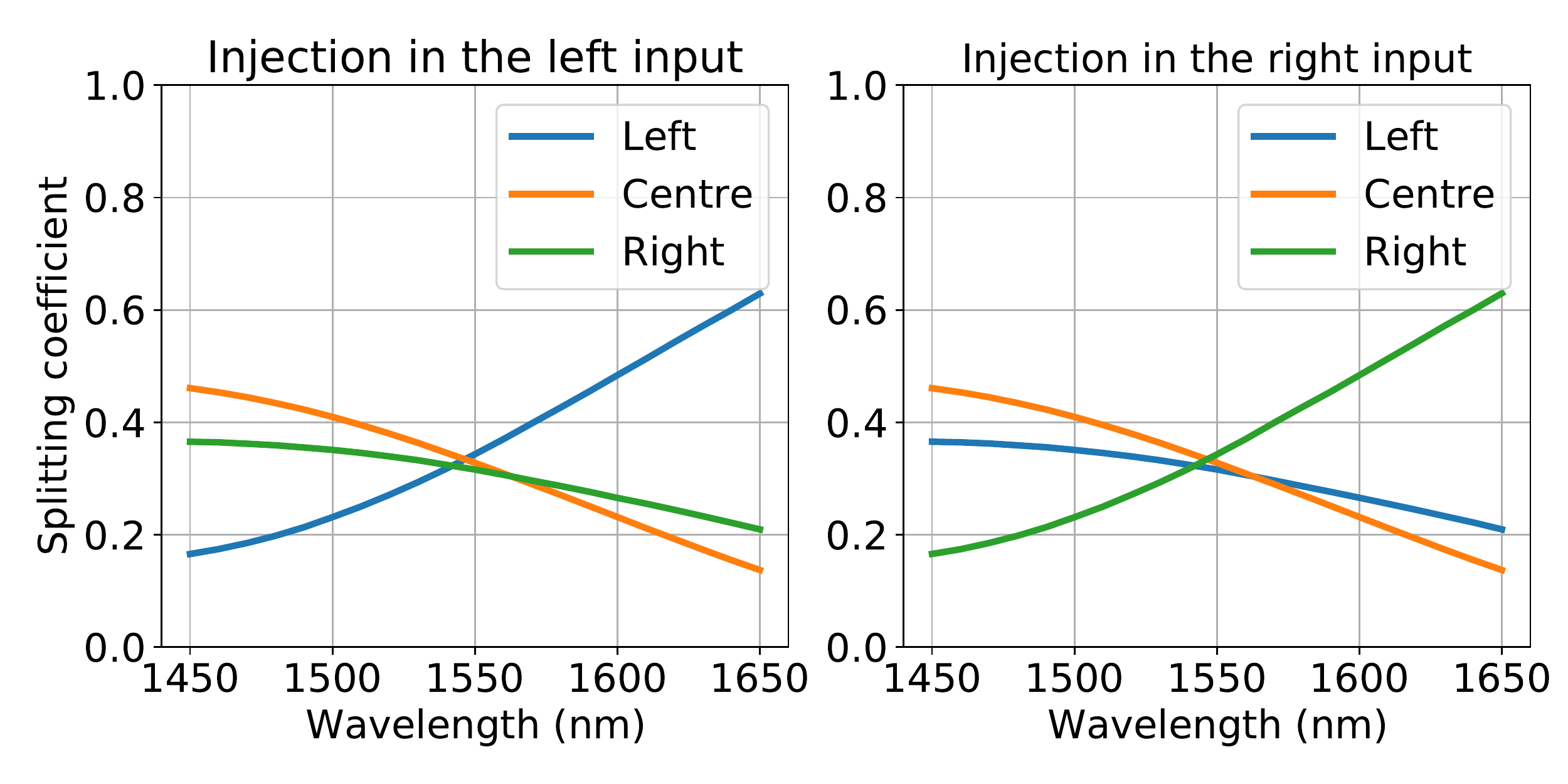}
    \caption{Splitting ratios between the three outputs for the injection of light into the left (left panel) or into the right input (right panel). The sum of the intensities from the three outputs is normalised to 1 for each wavelength channel.}
    \label{fig:tricoupler_splitting}
\end{figure}

As a first step in the numerical characterisation of the candidate, beams propagating from either the left or the right input through the tricoupler are simulated for different wavelengths to calculate the splitting ratios between the three outputs.
Figure~\ref{fig:tricoupler_splitting} shows these ratios to be wavelength-dependent but critically the centre waveguide ratios are {\it identical} between left and right inputs.
This correspondence ensures that while splitting ratios may be chromatic, the null remains entirely achromatic.
Furthermore, the chromaticity of the left and right outputs does not prevent the operation of the fringe tracking capability.
The response is incorporated in the experimental calibration of the fringe-tracker as a function of input phase delay.

\begin{figure*}
    \centering
    \begin{tabular}{c}
        \includegraphics[width=0.75\textwidth]{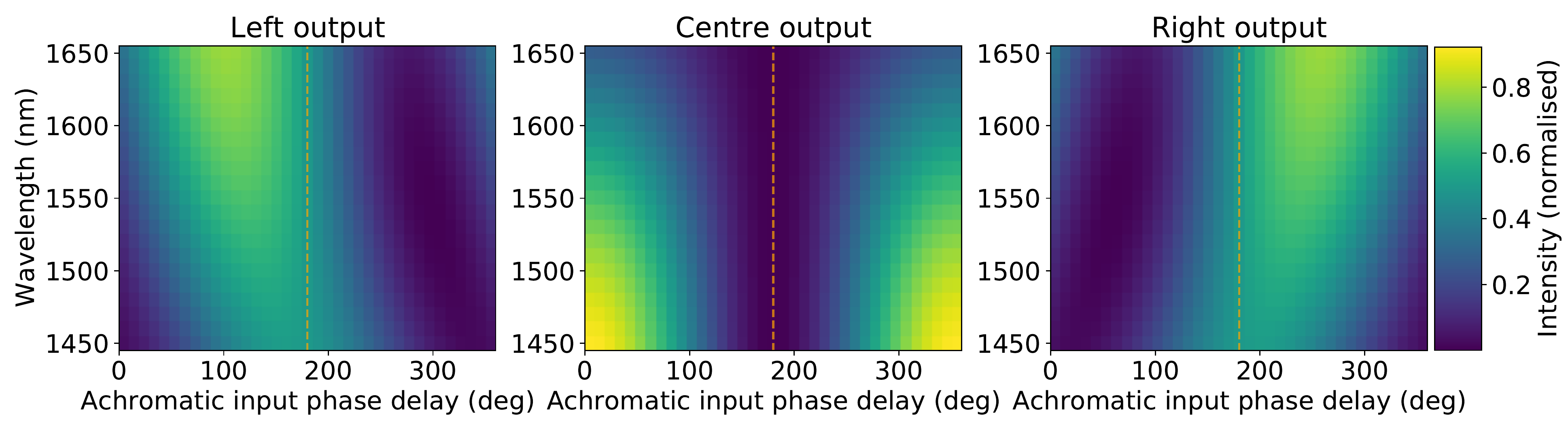}\\
        \includegraphics[width=0.75\textwidth]{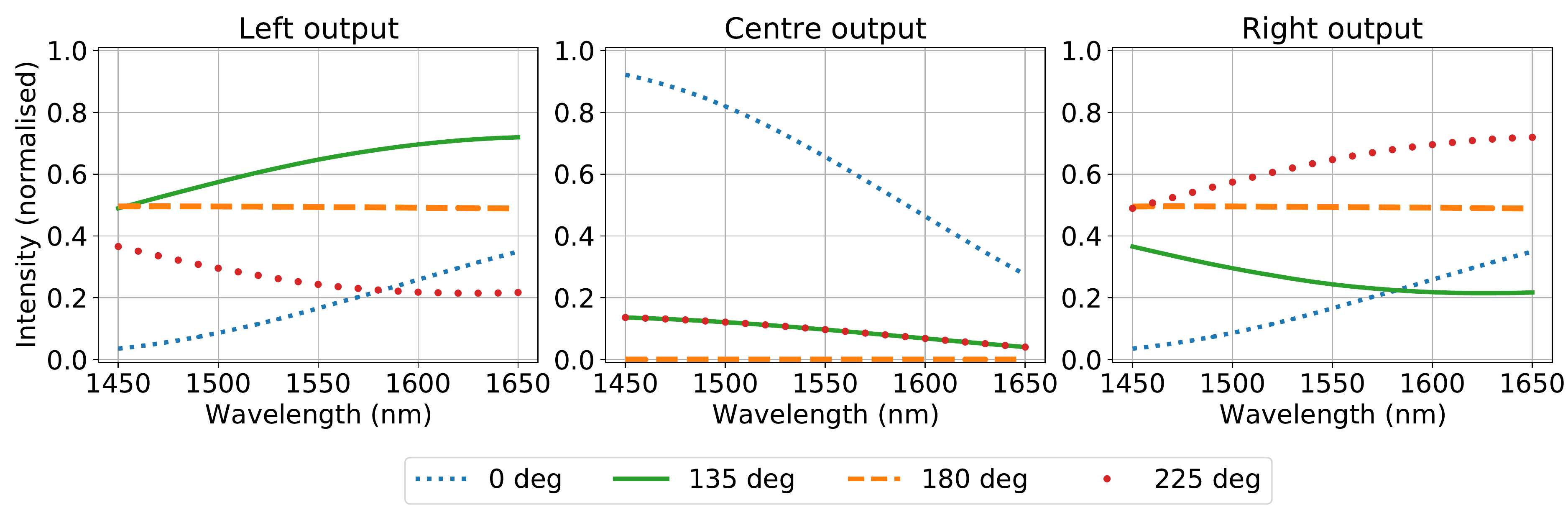}
    \end{tabular}
    \caption{Top: Map of intensities in the three outputs of the equilateral tricoupler (respectively the left, centre and right panel) versus the wavelength and the achromatic input phase delay between the incoming beams. The orange dashed line highlights the flux of the outputs in the case the beams have a 180$^\circ$ shift. The sum of the intensities from the three outputs is normalised to 1 for each wavelength channel.
    Bottom: Plots of the intensity with respect to wavelength for the three outputs for different phase shifts.}
    \label{fig:tricoupler_colormap}
\end{figure*}

Next the interferometric combination of two beams from the left and right tricoupler inputs was simulated, for several achromatic input phase delays, over a range of different wavelengths.
Results given in Figure~\ref{fig:tricoupler_colormap} show that the centre output is completely extinguished, regardless of the wavelength, for a phase delay of 180$^{\circ}$ -- corresponding to an achromatic destructive interference.
This achromatic behaviour, confirming our simple arguments based on symmetry above, is also present in the left and right outputs for the same phase difference.
For phase differences other than 180$^\circ$, all coupler outputs exhibit chromaticity.
It is also interesting to note that the plots of left and right outputs are mirror images of each other (Fig.~\ref{fig:tricoupler_colormap}, top).
Therefore when the center is acting as a null (the 180$^\circ$ achromatic input phase delay), the flux in the left and right must balance with 50\% each (Fig.~\ref{fig:tricoupler_colormap}, bottom). 
The strong intensity gradients with respect to phase in the Left and Right channels coincident with the null condition produces the ideal sensor for the recovery of the phase for fringe tracking.

\begin{figure*}
    \centering
    \begin{tabular}{c}
        \includegraphics[width=0.45\textwidth]{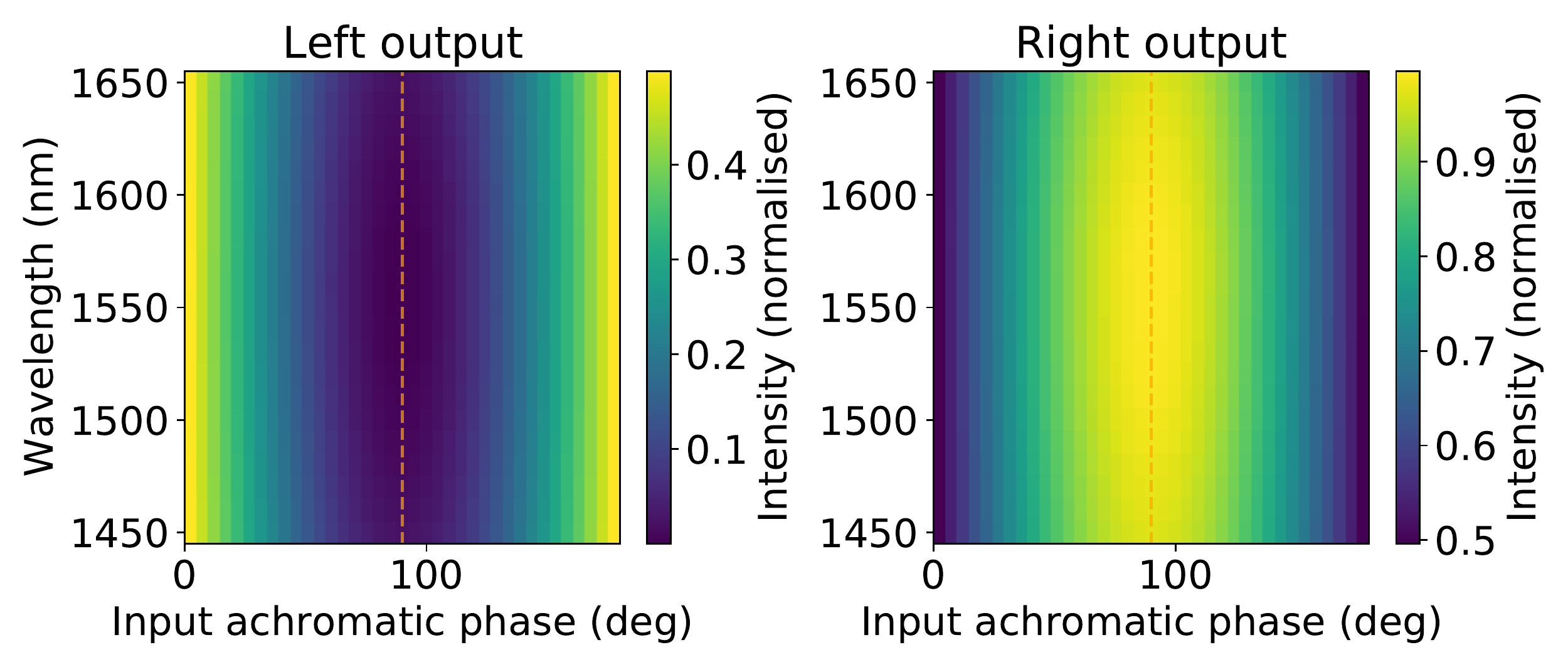}\\
        \includegraphics[width=0.45\textwidth]{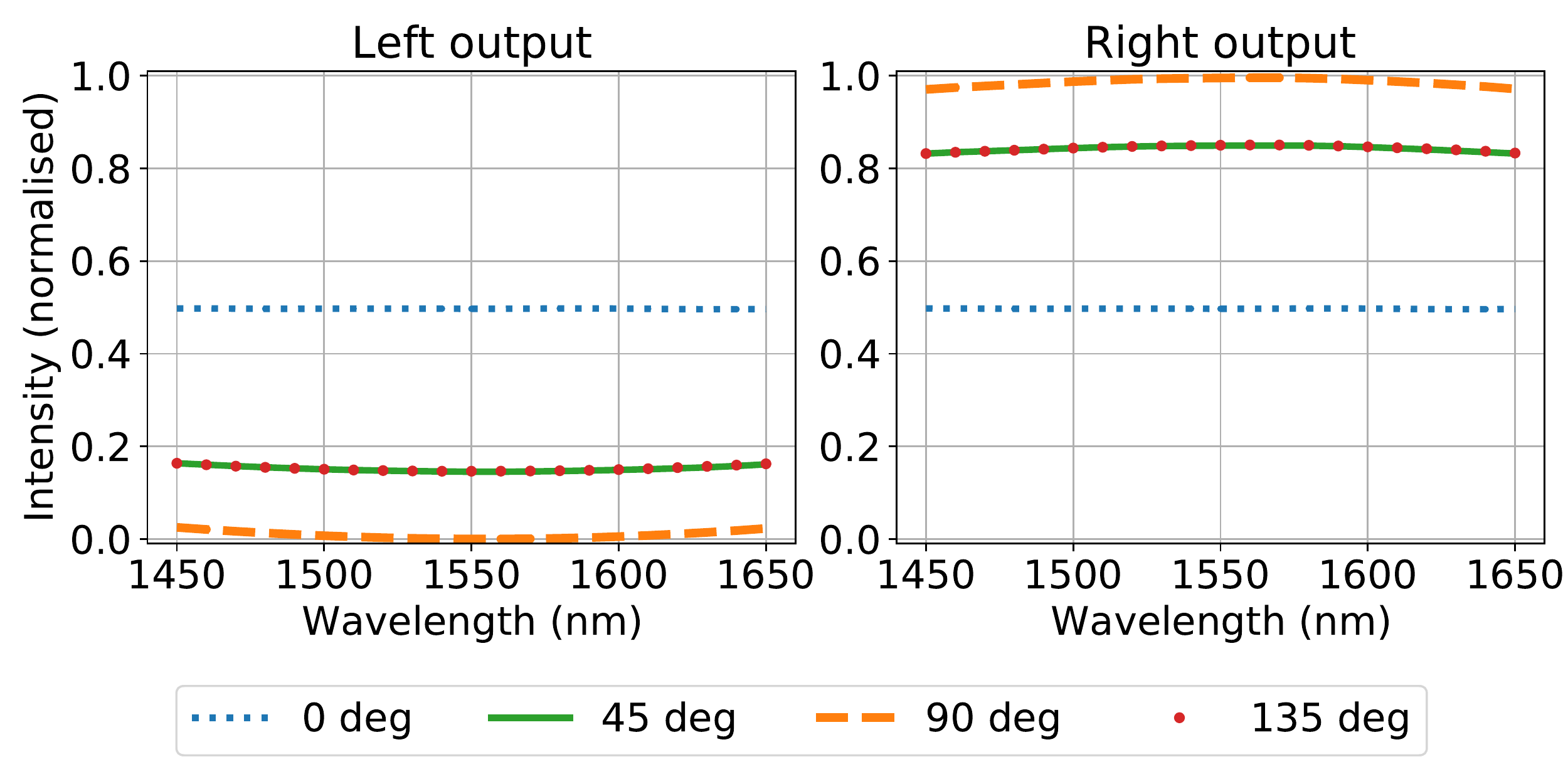}
    \end{tabular}
    \caption{Top: Map of intensities in the two outputs of the directional coupler (respectively the left and right panels) versus the wavelength and the achromatic input phase delay between the incoming beams. 
    The null condition is now obtained when the beams have a 90$^\circ$ phase shift, at which point the left output delivers the nulled channel while the right collects the constructive interference. 
    The orange dashed line highlights the location of this nulled operation. As before, the sum of Left and Right channels is normalised to 1 for each wavelength.
    Bottom: Plots of the intensity with respect to wavelength for the two outputs for different phase shifts.}
    \label{fig:cocoupler_colormap}
\end{figure*}

For comparison, an example $2\times2$ directional coupler (as presently implemented in GLINT) has been simulated using the same coupling length and refractive index as the tricoupler.
Figure~\ref{fig:cocoupler_colormap} depicts the way the light is shared between the two outputs as both the wavelength and the achromatic phase delay of the two incoming beams is varied.
The nulled signal (left output) clearly exhibits a chromatic behaviour for any achromatic input phase delay -- particularly for the phase difference of 90$^\circ$, where the beams are supposed to destructively interfere.\\

The simulations have therefore confirmed the earlier arguments for the superiority of the tricoupler over the directional coupler. 
The tricoupler provides a true achromatic null channel and strong flux gradients, of opposite signs, with input phase in the outer channel outputs demonstrate the capability to recover the phase between the input beams.
A quantification of this latter point, and the expected impact of the fringe tracking signal on a working nulling interferometer, is further explored in the sections below.

\section{Performance of a tricoupler on a nulling interferometer}
\label{sec:tricoupler_on_nulling}
This section aims to quantify the improvement of the light suppression of a single unresolved star -- quantified by the null depth measurement -- with the addition of fringe tracking enabled by the tricoupler described above, on top of the correction delivered by the AO system.
This is compared to the case of the present GLINT chip which uses a $2 \times 2$ directional coupler with no additional fringe tracking but with the correction delivered by the AO system.
After setting up the mathematical formalism, studies are presented spanning two regimes: {\it AO-residuals-limited} and {\it faint-target}.
The former is relevant to bright targets so that phase aberrations uncorrected by the AO system limit the performance of the nuller; such as tip-tilt effect due to vibrations of the telescope \cite{lozi2018} or the so-called ``low-wind effect'' (sharp phase discontinuities across the telescope's spiders believed due to thermal effects \cite{vievard2020}). 
The latter models the case where photon noise of the star is the limiting factor.
Therefore, the number of photons from the simulated target is 4800 times the photon noise for the AO-residuals-limited regime and 3 times the photon noise for the faint-target regime.

\subsection{Mathematical formalism}
\begin{figure}
    \centering
    \includegraphics[width=0.38\textwidth]{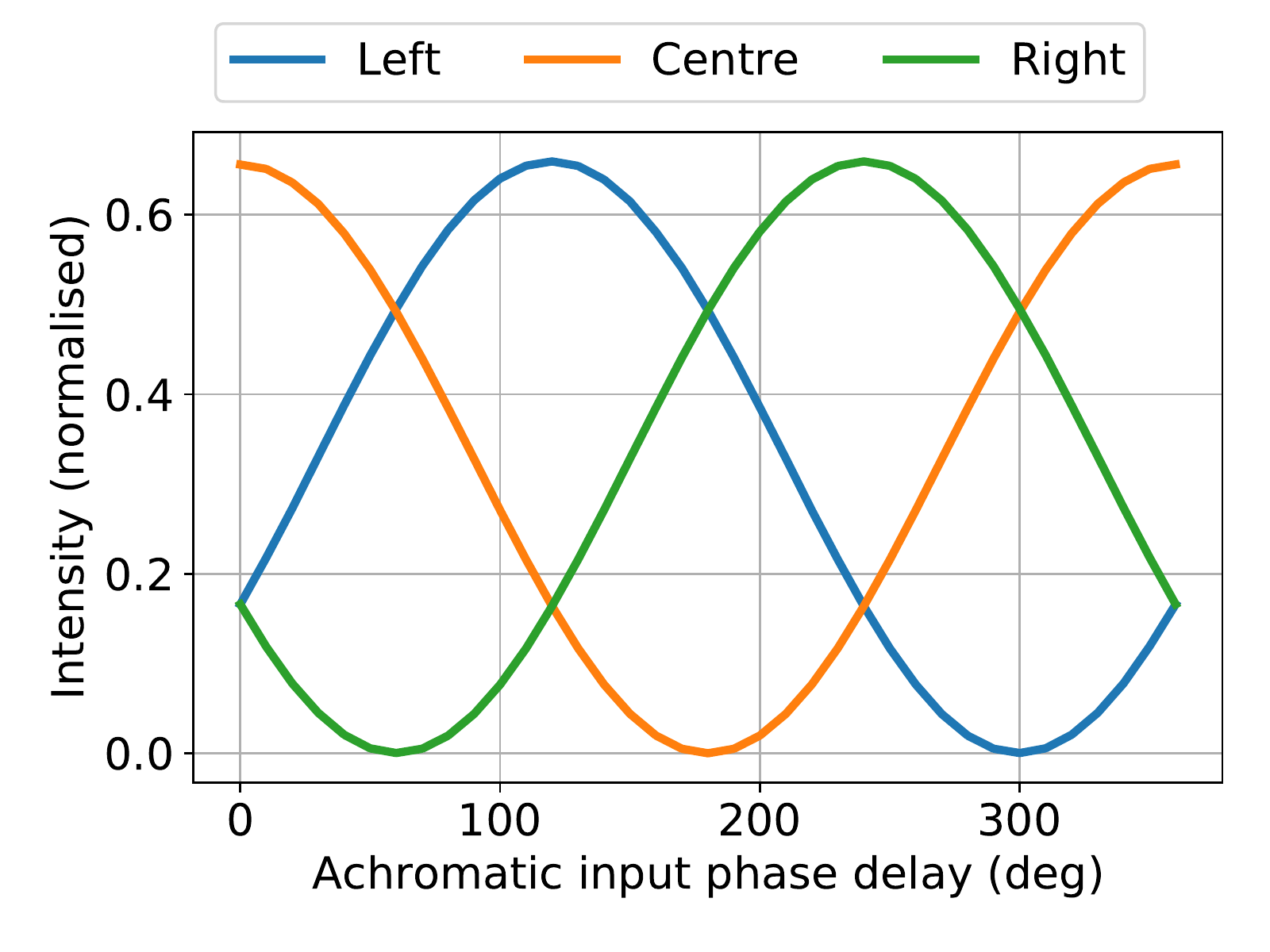}
    \caption{Normalised fluxes of the three outputs of the equilateral tricoupler with respect to the achromatic input phase delay between the incoming beams.}
    \label{fig:fringe_tracking_curve}
\end{figure}

The transfer matrix of the tricoupler \cite{xie2012} is 
\begin{equation}
    \label{eq:tricoupler}
    T_{\mathrm{tricoupler}} = \frac{1}{\sqrt{3}}
    \begin{pmatrix}
        1 & e^{j\frac{2\pi}{3}} & e^{j\frac{2\pi}{3}} \\
        e^{j\frac{2\pi}{3}} & 1 & e^{j\frac{2\pi}{3}}\\
        e^{j\frac{2\pi}{3}} & e^{j\frac{2\pi}{3}} & 1
    \end{pmatrix}
\end{equation}
with $j$ the complex number such that $j^2 = -1$.
This matrix considers an achromatic phase shift of $120^\circ$ between each waveguide so that the three outputs have balanced flux depending on the phase delay in the input (Fig.~\ref{fig:fringe_tracking_curve}).

Therefore, the intensities of the output fluxes are given by the matrix multiplication:
\begin{equation}
    \label{eq:intensities_tricoupler}
    \mathbf{I_{\mathrm{out}}} = 
    \left| T_{\mathrm{tricoupler}} \times \mathbf{a_{\mathrm{in}}}
    \right|^2
\end{equation}
with $\mathbf{I_{\mathrm{out}}}$ the 3-element vector of the output intensities (left, centre, right), $\mathbf{a_{\mathrm{in}}}$ is the 3-element vector of the incoming complex wavefront in the left and right waveguide (no wavefront is injected in the centre one).
$a_{\mathrm{in, left/right}}$ is such that:
\begin{equation}
    a_{\mathrm{in, left/right}} \propto e^{j\phi_{\mathrm{left/right}}}
\end{equation}
with $\phi_{\mathrm{left/right}}$ is the phase of the wavefront $a_{\mathrm{in, left/right}}$.
Therefore it follows that the intensities in the 3 outputs have the following expression:
\begin{multline}
    \label{eq:i_out_left}
    I_{\mathrm{out, left}}(\lambda) = \frac{1}{3}(I_{\mathrm{in, left}}(\lambda) + I_{\mathrm{in, right}}(\lambda)) + \\
    \frac{2}{3} \sqrt{I_{\mathrm{in, left}}(\lambda) I_{\mathrm{in, right}}(\lambda)} V \cos \left( \Delta \phi_{\mathrm{left}-\mathrm{right}} - \frac{2 \pi}{3} + \psi \right)
\end{multline}
\begin{multline}
    \label{eq:i_out_centre}
    I_{\mathrm{out, centre}}(\lambda) = \frac{1}{3}(I_{\mathrm{in, left}}(\lambda) + I_{\mathrm{in, right}}(\lambda)) + \\
    \frac{2}{3} \sqrt{I_{\mathrm{in, left}}(\lambda) I_{\mathrm{in, right}}(\lambda)} V \cos \left( \Delta \phi_{\mathrm{left}-\mathrm{right}} + \psi \right)
\end{multline}
\begin{multline}
    \label{eq:i_out_right}
    I_{\mathrm{out, right}}(\lambda) = \frac{1}{3}(I_{\mathrm{in, left}}(\lambda) + I_{\mathrm{in, right}}(\lambda)) + \\
    \frac{2}{3} \sqrt{I_{\mathrm{in, left}}(\lambda) I_{\mathrm{in, right}}(\lambda)} V \cos \left( \Delta \phi_{\mathrm{left}-\mathrm{right}} + \frac{2 \pi}{3} + \psi \right)
\end{multline}
with $I_{\mathrm{in, left}} = |a_{\mathrm{in, left}}|^2$ and $I_{\mathrm{in, right}} = |a_{\mathrm{in, right}}|^2$ which depend on the wavelength $\lambda$, $V$ and $\psi$ are respectively the modulus and the phase of the complex degree of spatial coherence of the object, and $\Delta \phi_{\mathrm{left}-\mathrm{right}} = \phi_{\mathrm{left}} - \phi_{\mathrm{right}}$ the phase difference between the left and right wavefronts.
For the sake of clarity, it is assumed henceforth that the source is fully coherent: $V=1$ and $\psi=0$.

The null depth is linked to $V$ by $N = \frac{1-V}{1+V}$.
It is also defined as the ratio of the intensities of the destructive (\eqref{eq:i_out_centre}) to constructive interference (obtained by a linear combination of \eqref{eq:i_out_left}, \eqref{eq:i_out_centre} and \eqref{eq:i_out_right}):
\begin{multline}
    \label{eq:tricoupler_null}
    N_{\mathrm{tricoupler}}(\lambda) = \\
    \frac{I_{\mathrm{out, centre}}(\lambda)}{2/3 (I_{\mathrm{out, left}}(\lambda) + I_{\mathrm{out, right}}(\lambda)) - 1/3 I_{\mathrm{out, centre}}(\lambda)},
\end{multline}
where $I_{\mathrm{out, left}}$, $I_{\mathrm{out, centre}}$ and $I_{\mathrm{out, right}}$ are respectively the intensities measured in the left, centre and right output (Fig.\ref{fig:tricoupler_scheme}).

The phase between the left and right wavefronts is recovered with:
\begin{multline}
    \label{eq:delta_phi}
    \Delta \phi_{\mathrm{left}-\mathrm{right}} = \\
    \sqrt{3} \arctan \left( \frac{I_{\mathrm{out, left}}(\lambda) - I_{\mathrm{out, right}}(\lambda)}{I_{\mathrm{out, left}}(\lambda) + I_{\mathrm{out, right}}(\lambda) - 2 I_{\mathrm{out, centre}}(\lambda)} \right)
\end{multline}
and the differential piston $\delta_{\mathrm{left}-\mathrm{right}}$ is defined by:
\begin{equation}
    \label{eq:delta_opd}
    \delta_{\mathrm{left}-\mathrm{right}} = \frac{\lambda}{2 \pi} \Delta \phi.
\end{equation}
The spectral dispersion allows the extension of the range of validity for \eqref{eq:delta_opd} beyond $\pm \frac{\lambda}{2}$, by applying the method described by Choquet~\textit{et~al.} \cite{choquet2014}.\\

For comparison with the existing GLINT coupler (result in Section~\ref{sec:simulations} below), an ideal achromatic directional coupler is considered to operate according to the following transfer matrix:
\begin{equation}
    T_{\mathrm{di-coupler}} = \frac{1}{\sqrt{2}}
    \begin{pmatrix}
        1 & e^{-j\frac{\pi}{2}} \\
        e^{-j\frac{\pi}{2}} & 1
    \end{pmatrix}
    .
\end{equation}
The null depth is defined as:
\begin{equation}
    \label{eq:cocoupler}
    N_{\mathrm{di-coupler}}(\lambda) = \frac{I_{\mathrm{out, left}}(\lambda)}{I_{\mathrm{out, right}}(\lambda)}.
\end{equation}

\subsection{Simulations of the nuller with both photonic devices}
\label{sec:simulations}
A nulling interferometer, equipped with a tricoupler or a directional coupler, has been simulated.
This entails interferometric beam combination of two wavefronts extracted as sub-apertures from a larger 8\,m pupil that has undergone partial correction by an adaptive optics system.
These two sub-pupils are then phase-shifted relative to each other assuming a chromatic free-space delay (as is done for GLINT -- not the perfect achromatic phase delays discussed earlier).
The instrument is simulated to have the same spectroscope as GLINT, delivering dispersed light from the output channels spanning the range 1450--1650~nm with spectral bins of 5~nm \cite{martinod2021}.\\

The simulated wavefronts are generated by a Kolmogorov phase screen with Fried parameter, $r_0=16$~cm\footnote{\url{https://www.naoj.org/Observing/Telescope/ImageQuality/Seeing/}}, and coherence time, $\tau_0=5.14$~ms \cite{travouillon2009}, corresponding to median conditions at Mauna Kea at 500~nm.
The performance of the AO is tuned so that the standard deviation of the differential piston after correction is 163~nm, equivalent to a Strehl ratio of 64\% at 1.55~$\mu$m, corresponding to what is observed in real on-sky GLINT data \cite{martinod2021}.
The wavefronts are then cropped into two sub-pupils of 1~m of diameter which are separated by 5.55~m (to the scale of the simulation) on the full 8\,m pupil.
The differential piston is defined by the difference between the averages of the phase between the two sub-pupils.
Atmospheric scintillation is not incorporated into the model.
The spatial filtering effect of the waveguide is modelled by the injection efficiency, $\rho = 0.8 Sr$\cite{coude2000}, applied to the intensity of the wavefront, with the Strehl ratio, $Sr$, calculated with the Marechal approximation: $Sr = e^{-\sigma^2}$ where $\sigma$ is the RMS of the phase across the sub-pupil.
The phase delay between the incoming beams is tuned by chromatic air-delay lines which consist of mobile segmented mirrors similar to the GLINT's MEMS \cite{norris2020}.

The resulting wavefronts are combined following \eqref{eq:intensities_tricoupler} and the intensities of the outputs are collected to measure the null depth and the differential piston.
A simulated fringe tracker measures the incoming differential piston with \eqref{eq:delta_phi} and \eqref{eq:delta_opd} and changes the positions of the segmented mirrors to compensate for the measured piston.
The fringe tracker is modelled as a basic closed integrator servo control loop with a total latency of 1~ms (a typical value for wavefront control \cite{currie2020}), including an exposure time of 0.5~ms.
This controller would be straightforward to implement on GLINT with existing hardware and software.
In a nutshell, an integrator servo controller loop iteratively measures the difference between the chromatic input phase delay, measured as a differential piston and the reference (here 180$^\circ$ at 1.55~$\mu$m).
Then, it applies a correction $c$ to keep the error the closest to zero so that
\begin{equation}
    c_k = K_{\mathrm{I}} \sum_{i=1}^{k-1} \delta_i - \frac{\lambda_0}{2},    
\end{equation}
with $c_k$ the correction for the k-th iteration, $K_{\mathrm{I}}$ the gain of the integrator loop, $\delta_i$ the input differential piston measured at the iteration $i < k$ and $\frac{\lambda_0}{2}$ the reference.
The servo-loop is optimized when the statistical dispersion of the null depth is minimized.
Values used are $K_{\mathrm{I}} = 1$ in the AO-residuals-limited regime and $K_{\mathrm{I}} = 0.05$ in the faint-target regime.\\

\begin{figure}
    \centering
    \begin{tabular}{c}
         \includegraphics[width=0.4\textwidth]{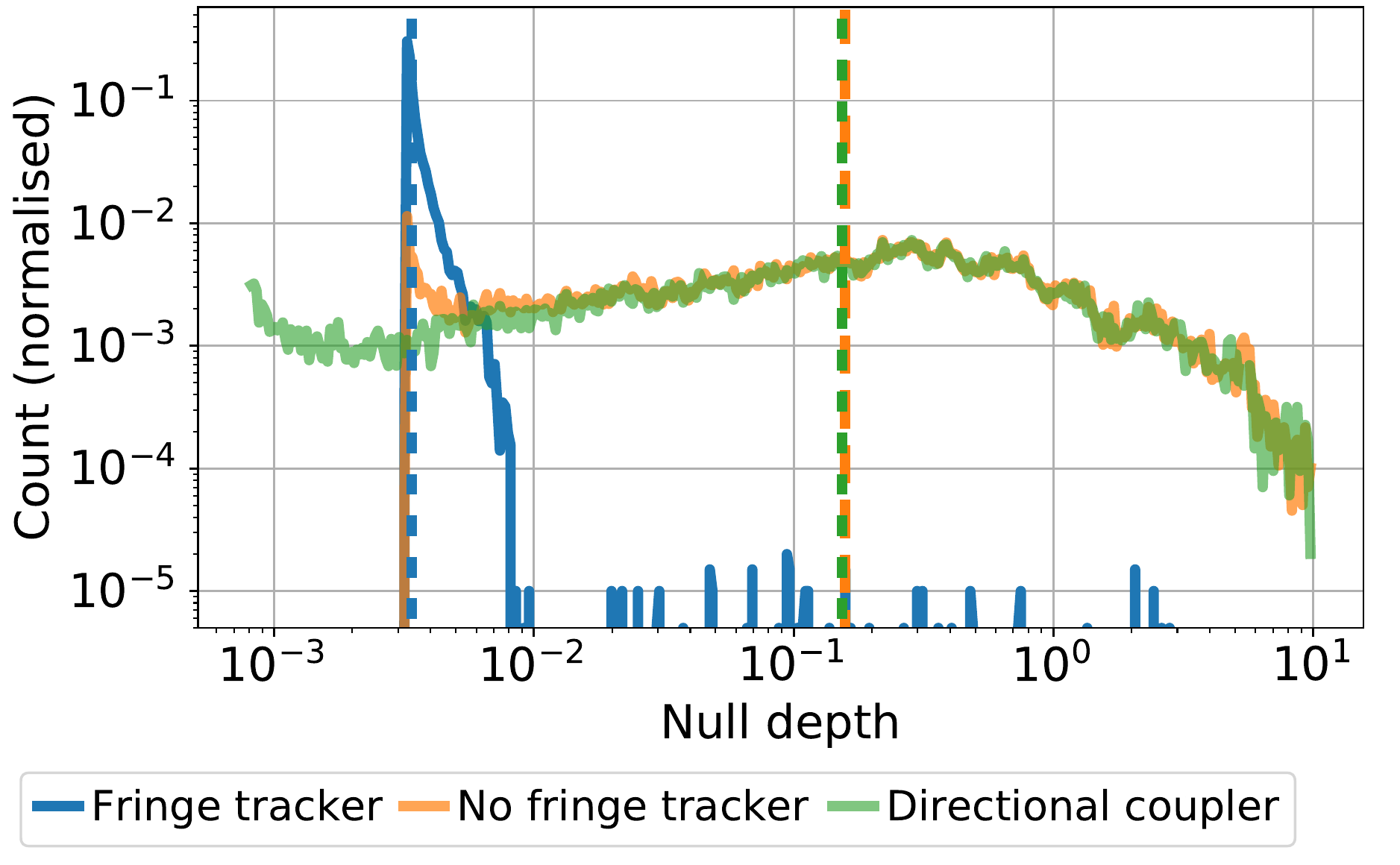}\\
         \includegraphics[width=0.4\textwidth]{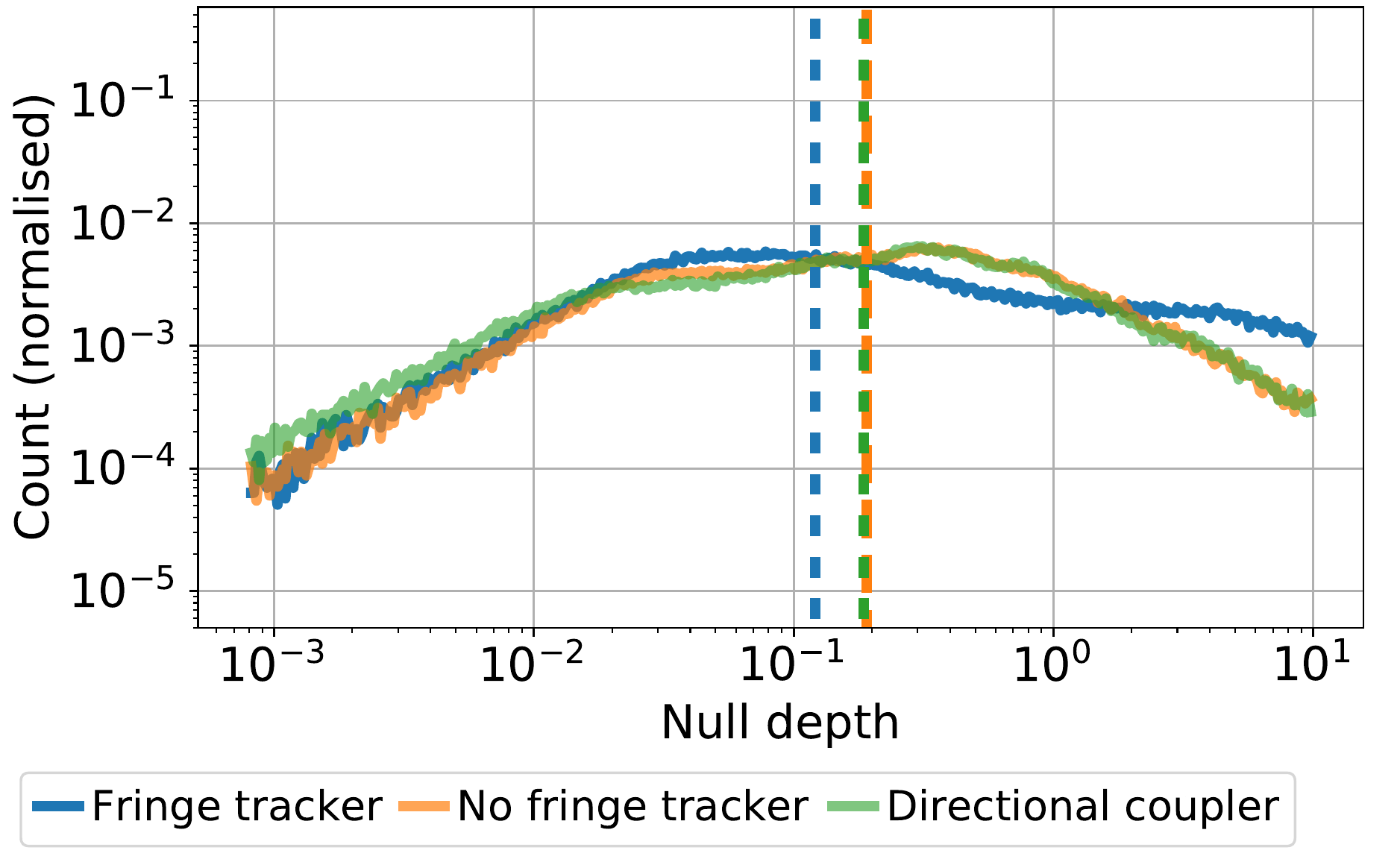}
    \end{tabular}
    \caption{Histograms of the null depth for the cases of tricoupler with fringe tracking (solid blue curve), without fringe tracking (solid orange curve) and the directional coupler (solid green curve) in the {\it AO-residuals-limited} (top panel) and {\it faint-target} (bottom panel) regimes. The blue, orange and green dashed lines respectively represent the median of the distributions of the null depths in the three configurations. The orange and green dashed lines are almost overlapping due to similar performance between the cases of the directional coupler and the tricoupler with no fringe tracking operation.}
    \label{fig:histo}
\end{figure}

A time-sequence of 200\,000 data points was simulated in order to provide sufficient samples for the statistical analysis in the following section.
A similar sequence, with the same atmospheric phase screen, was generated with the directional coupler.
The simulations were run in the AO-residuals-limited regime and also in the faint-target regime.
Three different instrumental configurations were trialed: (i) using the tricoupler with the fringe tracker operational (ii) using the tricoupler but turning fringe tracking off, and (iii) using the existing directional coupler.
The flux of the outputs were averaged over the bandwidth then the null depths were calculated with \eqref{eq:tricoupler} for the cases (i) and (ii), and \eqref{eq:cocoupler} for the case (iii).
The statistical distributions of the null depths are then compiled, spanning these 3 configurations over two noise regimes (Fig.~\ref{fig:histo} and Tab.~\ref{tab:results}).

\begin{table}
    \centering
    \caption{Median value and standard deviation of the null depth distributions for the three configurations in the two regimes. \textit{FT} means \textit{fringe-tracker}.}
    \begin{tabular}{lcc}
         Configuration & Median & Standard deviation \\
         \hline
         \hline
         \multicolumn{3}{c}{AO-residuals-limited regime}\\
         \hline
         Tricoupler + FT & $3.3 \times 10^{-3}$ & $1.5 \times 10^{-2}$\\
         Tricoupler & $0.15$ & 0.91\\
         Directional coupler & $0.15$ & 0.9\\
         \hline
         \multicolumn{3}{c}{Faint-target regime}\\
         \hline
         Tricoupler + FT & $0.11$ & 1.6\\
         Tricoupler & $0.18$ & 1.0\\
         Directional coupler & $0.18$ & 1.0
    \end{tabular}
    \label{tab:results}
\end{table}

\section{Discussion}
\label{sec:discussion}
Figure~\ref{fig:histo} immediately illustrates that the use of fringe tracking dramatically improves the statistics of the null depth in AO-residuals-limited regime: the contrast is deeper and determined with less dispersion (Tab.~\ref{tab:results}).
The use of the tricoupler, which enables simultaneous fringe tracking and null depth measurements, reaches a null depth 45~times better than with a directional coupler in this regime.
On the other hand, when the fringe-tracking operation is disabled, the performance of the tricoupler is the same as that of the directional coupler.
This finding is explained by the fact that the simple simulations performed here did not incorporate the chromatic behaviour of the components under study.
Under conditions where a truly achromatic phase-difference could be maintained upon injection, then indeed the intrinsically achromatic null provided by the tricoupler should confer advantages compared to the chromatic behavior of the directional coupler.
A more sophisticated treatment of chromaticity and phase is presently being formulated to represent this added complexity.

In the faint-target regime, the median values for the three configurations are equivalent, and the precision of the configuration with the fringe-tracking tricoupler is worse than the two others.
This result is expected because the signal is dominated by the photon and readout noise; the fringe tracker is unable to provide meaningful correction to the input phase delay and merely adds noise, enlarging the distribution of the null depth.

Finally, up to two orders of magnitude gain in null depth sensitivity can be anticipated with the NSC method \cite{hanot2011, Mennesson2011, kuhn2015, serabyn2019, martinod2021}, on top of the gain obtained with the tricoupler and the fringe-tracking.
This places the GLINT instrument into a regime of a null depth of around $10^{-5}$ which makes the detection of faint companions, such as exoplanets close to their host star, readily accessible.

\section{Conclusion}
The photonic tricoupler efficiently combines, in a single device, functionality to perform both fringe tracking and nulling interferometry.
It delivers an achromatic null enforced by symmetry considerations. 
Instantaneous measurements of the phase of the incoming beams can be made without non-common path error, and these allow for fast stabilisation of the null channel where an active element is available to perform the compensation.
In addition, the fringe-tracking capability allows the integration of this photonic device into an AO system, improving wavefront correction while simultaneously delivering an astrophysical null-channel science.
The ultrafast laser inscription technique provides an immediate pathway to the fabrication of equilateral-triangle tricoupler components with advantageous symmetry and tunable coupling parameters.
The simulation of the impact of implementation of an achromatic tricoupler within a working nulling interferometer allowed us to quantify the expected gains in performance.
An improvement of the null depth of the GLINT instrument by a factor 45 in the AO-residuals-limited regime was seen, assuming median atmospheric conditions at Mauna Kea.
Statistical data analysis by the established NSC method was shown to bring the null depth to $10^{-5}$: a level of performance that brings the detection of exoplanets close to their host star within reach.

\bibliographystyle{unsrtnat}
\bibliography{biblio}  

\begin{thebibliography}{44}
\providecommand{\natexlab}[1]{#1}
\providecommand{\url}[1]{\texttt{#1}}
\expandafter\ifx\csname urlstyle\endcsname\relax
  \providecommand{\doi}[1]{doi: #1}\else
  \providecommand{\doi}{doi: \begingroup \urlstyle{rm}\Url}\fi

\bibitem[{Martinod} et~al.(2018){Martinod}, {Mourard}, {B{\'e}rio}, {Perraut},
  {Meilland}, {Bailet}, {Bresson}, {ten Brummelaar}, {Clausse}, {Dejonghe},
  {Ireland}, {Millour}, {Monnier}, {Sturmann}, {Sturmann}, and
  {Tallon}]{martinod2018}
M.~A. {Martinod}, D.~{Mourard}, P.~{B{\'e}rio}, K.~{Perraut}, A.~{Meilland},
  C.~{Bailet}, Y.~{Bresson}, T.~{ten Brummelaar}, J.~M. {Clausse},
  J.~{Dejonghe}, M.~{Ireland}, F.~{Millour}, J.~D. {Monnier}, J.~{Sturmann},
  L.~{Sturmann}, and M.~{Tallon}.
\newblock {Fibered visible interferometry and adaptive optics: FRIEND at
  CHARA}.
\newblock \emph{Astron. Astrophys}, 618:\penalty0 A153, October 2018.
\newblock \doi{10.1051/0004-6361/201731386}.

\bibitem[{Huby} et~al.(2012){Huby}, {Perrin}, {Marchis}, {Lacour}, {Kotani},
  {Duch{\^e}ne}, {Choquet}, {Gates}, {Woillez}, {Lai}, {F{\'e}dou}, {Collin},
  {Chapron}, {Arslanyan}, and {Burns}]{huby2012}
E.~{Huby}, G.~{Perrin}, F.~{Marchis}, S.~{Lacour}, T.~{Kotani},
  G.~{Duch{\^e}ne}, E.~{Choquet}, E.~L. {Gates}, J.~M. {Woillez}, O.~{Lai},
  P.~{F{\'e}dou}, C.~{Collin}, F.~{Chapron}, V.~{Arslanyan}, and K.~J. {Burns}.
\newblock {FIRST, a fibered aperture masking instrument. I. First on-sky test
  results}.
\newblock \emph{Astron. Astrophys}, 541:\penalty0 A55, May 2012.
\newblock \doi{10.1051/0004-6361/201118517}.

\bibitem[{Gravity Collaboration}(2017)]{gravity2017}
{Gravity Collaboration}.
\newblock {First light for GRAVITY: Phase referencing optical interferometry
  for the Very Large Telescope Interferometer}.
\newblock \emph{Astron. Astrophys}, 602:\penalty0 A94, June 2017.
\newblock \doi{10.1051/0004-6361/201730838}.

\bibitem[{Gravity Collaboration}(2019)]{2019A&A...623L..11G}
{Gravity Collaboration}.
\newblock {First direct detection of an exoplanet by optical interferometry.
  Astrometry and K-band spectroscopy of HR 8799 e}.
\newblock \emph{Astron. Astrophys}, 623:\penalty0 L11, March 2019.
\newblock \doi{10.1051/0004-6361/201935253}.

\bibitem[{Marois} et~al.(2008){Marois}, {Macintosh}, {Barman}, {Zuckerman},
  {Song}, {Patience}, {Lafreni{\`e}re}, and {Doyon}]{Marois2008}
Christian {Marois}, Bruce {Macintosh}, Travis {Barman}, B.~{Zuckerman}, Inseok
  {Song}, Jennifer {Patience}, David {Lafreni{\`e}re}, and Ren{\'e} {Doyon}.
\newblock {Direct Imaging of Multiple Planets Orbiting the Star HR 8799}.
\newblock \emph{Science}, 322\penalty0 (5906):\penalty0 1348, November 2008.
\newblock \doi{10.1126/science.1166585}.

\bibitem[{Schworer} and {Tuthill}(2015)]{Schworer2015}
Guillaume {Schworer} and Peter~G. {Tuthill}.
\newblock {Predicting exoplanet observability in time, contrast, separation,
  and polarization, in scattered light}.
\newblock \emph{Astron. Astrophys}, 578:\penalty0 A59, June 2015.
\newblock \doi{10.1051/0004-6361/201424202}.

\bibitem[{Bracewell}(1978)]{Bracewell1978}
R.~N. {Bracewell}.
\newblock {Detecting nonsolar planets by spinning infrared interferometer}.
\newblock \emph{Nature}, 274\penalty0 (5673):\penalty0 780--781, August 1978.
\newblock \doi{10.1038/274780a0}.

\bibitem[{Lagadec} et~al.(2018){Lagadec}, {Norris}, {Gross}, {Arriola},
  {Gretzinger}, {Cvetojevic}, {Lawrence}, {Withford}, and
  {Tuthill}]{lagadec2018}
Tiphaine {Lagadec}, Barnaby {Norris}, Simon {Gross}, Alexander {Arriola},
  Thomas {Gretzinger}, Nick {Cvetojevic}, Jon {Lawrence}, Michael {Withford},
  and Peter {Tuthill}.
\newblock {GLINT South: a photonic nulling interferometer pathfinder at the
  Anglo-Australian Telescope for high contrast imaging of substellar
  companions}.
\newblock In Michelle~J. {Creech-Eakman}, Peter~G. {Tuthill}, and Antoine
  {M{\'e}rand}, editors, \emph{Optical and Infrared Interferometry and Imaging
  VI}, volume 10701 of \emph{Proc. SPIE}, page 107010V, July 2018.
\newblock \doi{10.1117/12.2313171}.

\bibitem[{Norris} et~al.(2020){Norris}, {Cvetojevic}, {Lagadec}, {Jovanovic},
  {Gross}, {Arriola}, {Gretzinger}, {Martinod}, {Guyon}, {Lozi}, {Withford},
  {Lawrence}, and {Tuthill}]{norris2020}
Barnaby R.~M. {Norris}, Nick {Cvetojevic}, Tiphaine {Lagadec}, Nemanja
  {Jovanovic}, Simon {Gross}, Alexander {Arriola}, Thomas {Gretzinger},
  Marc-Antoine {Martinod}, Olivier {Guyon}, Julien {Lozi}, Michael~J.
  {Withford}, Jon~S. {Lawrence}, and Peter {Tuthill}.
\newblock {First on-sky demonstration of an integrated-photonic nulling
  interferometer: the GLINT instrument}.
\newblock \emph{Mon. Not. Roy. Astron. Soc.}, 491\penalty0 (3):\penalty0
  4180--4193, January 2020.
\newblock \doi{10.1093/mnras/stz3277}.

\bibitem[{Guyon} et~al.(2011){Guyon}, {Martinache}, {Clergeon}, {Russell},
  {Groff}, and {Garrel}]{Guyon2011}
Olivier {Guyon}, Frantz {Martinache}, Christophe {Clergeon}, Robert {Russell},
  Tyler {Groff}, and Vincent {Garrel}.
\newblock \emph{{Wavefront control with the Subaru Coronagraphic Extreme
  Adaptive Optics (SCExAO) system}}, volume 8149 of \emph{Proc. SPIE}, page
  814908.
\newblock SPIE, 2011.
\newblock \doi{10.1117/12.894293}.

\bibitem[{Jovanovic} et~al.(2013){Jovanovic}, {Guyon}, {Martinache},
  {Clergeon}, {Singh}, {Vievard}, {Kudo}, {Garrel}, {Norris}, {Tuthill},
  {Stewart}, {Huby}, {Perrin}, and {Lacour}]{Jovanovic2013}
Nemanja {Jovanovic}, Olivier {Guyon}, Frantz {Martinache}, Christophe
  {Clergeon}, Garima {Singh}, Sebastien {Vievard}, Tomoyuki {Kudo}, Vincent
  {Garrel}, Barnaby {Norris}, Peter {Tuthill}, Paul {Stewart}, Elsa {Huby}, Guy
  {Perrin}, and Sylvestre {Lacour}.
\newblock {SCExAO as a precursor to an ELT exoplanet direct imaging
  instrument}.
\newblock In \emph{Proceedings of the Third AO4ELT Conference}, page~94,
  December 2013.
\newblock \doi{10.12839/AO4ELT3.13396}.

\bibitem[{Jovanovic} et~al.(2015){Jovanovic}, {Martinache}, {Guyon},
  {Clergeon}, {Singh}, {Kudo}, {Garrel}, {Newman}, {Doughty}, {Lozi}, {Males},
  {Minowa}, {Hayano}, {Takato}, {Morino}, {Kuhn}, {Serabyn}, {Norris},
  {Tuthill}, {Schworer}, {Stewart}, {Close}, {Huby}, {Perrin}, {Lacour},
  {Gauchet}, {Vievard}, {Murakami}, {Oshiyama}, {Baba}, {Matsuo}, {Nishikawa},
  {Tamura}, {Lai}, {Marchis}, {Duchene}, {Kotani}, and
  {Woillez}]{Jovanovic2015}
N.~{Jovanovic}, F.~{Martinache}, O.~{Guyon}, C.~{Clergeon}, G.~{Singh},
  T.~{Kudo}, V.~{Garrel}, K.~{Newman}, D.~{Doughty}, J.~{Lozi}, J.~{Males},
  Y.~{Minowa}, Y.~{Hayano}, N.~{Takato}, J.~{Morino}, J.~{Kuhn}, E.~{Serabyn},
  B.~{Norris}, P.~{Tuthill}, G.~{Schworer}, P.~{Stewart}, L.~{Close},
  E.~{Huby}, G.~{Perrin}, S.~{Lacour}, L.~{Gauchet}, S.~{Vievard},
  N.~{Murakami}, F.~{Oshiyama}, N.~{Baba}, T.~{Matsuo}, J.~{Nishikawa},
  M.~{Tamura}, O.~{Lai}, F.~{Marchis}, G.~{Duchene}, T.~{Kotani}, and
  J.~{Woillez}.
\newblock {The Subaru Coronagraphic Extreme Adaptive Optics System: Enabling
  High-Contrast Imaging on Solar-System Scales}.
\newblock \emph{Publications of the Astronomical Society of the Pacific},
  127\penalty0 (955):\penalty0 890, September 2015.
\newblock \doi{10.1086/682989}.

\bibitem[{Hanot} et~al.(2011){Hanot}, {Mennesson}, {Martin}, {Liewer}, {Loya},
  {Mawet}, {Riaud}, {Absil}, and {Serabyn}]{hanot2011}
C.~{Hanot}, B.~{Mennesson}, S.~{Martin}, K.~{Liewer}, F.~{Loya}, D.~{Mawet},
  P.~{Riaud}, O.~{Absil}, and E.~{Serabyn}.
\newblock {Improving Interferometric Null Depth Measurements using Statistical
  Distributions: Theory and First Results with the Palomar Fiber Nuller}.
\newblock \emph{Astrophys. J}, 729\penalty0 (2):\penalty0 110, March 2011.
\newblock \doi{10.1088/0004-637X/729/2/110}.

\bibitem[{Martinod} et~al.(2021){Martinod}, {Norris}, {Tuthill}, {Lagadec},
  {Jovanovic}, {Cvetojevic}, Gross, {Arriola}, {Gretzinger}, {Withford},
  {Guyon}, {Lozi}, {Lawrence}, and {Leon-Saval}]{martinod2021}
M-.A. {Martinod}, B.~{Norris}, P.~{Tuthill}, T.~{Lagadec}, N.~{Jovanovic},
  N.~{Cvetojevic}, S.~Gross, A.~{Arriola}, T.~{Gretzinger}, M.~J. {Withford},
  O.~{Guyon}, J.~{Lozi}, J.~S. {Lawrence}, and S.~{Leon-Saval}.
\newblock {Scalable photonic-based nulling interferometry with the dispersed,
  multi-baseline GLINT instrument}.
\newblock \emph{Nature Communications}, 12:\penalty0 2465, April 2021.
\newblock \doi{10.1038/s41467-021-22769-x}.

\bibitem[Martin et~al.(2014)Martin, Heidmann, Rauch, Jocou, and
  Courjal]{Martin2014}
Guillermo Martin, Samuel Heidmann, Jean-Yves Rauch, Laurent Jocou, and Nadège
  Courjal.
\newblock {Electro-optic fringe locking and photometric tuning using a
  two-stage Mach–Zehnder lithium niobate waveguide for high-contrast
  mid-infrared interferometry}.
\newblock \emph{Optical Engineering}, 53\penalty0 (3):\penalty0 1 -- 8, 2014.
\newblock \doi{10.1117/1.OE.53.3.034101}.
\newblock URL \url{https://doi.org/10.1117/1.OE.53.3.034101}.

\bibitem[{Weber} et~al.(2004){Weber}, {Barillot}, {Haguenauer}, {Kern},
  {Schanen-Duport}, {Labeye}, {Pujol}, and {Sodnik}]{weber2004}
Valerie {Weber}, Marc {Barillot}, Pierre {Haguenauer}, Pierre~Y. {Kern},
  Isabelle {Schanen-Duport}, Pierre~R. {Labeye}, Laurence {Pujol}, and Zoran
  {Sodnik}.
\newblock {Nulling interferometer based on an integrated optics combiner}.
\newblock In Wesley~A. {Traub}, editor, \emph{New Frontiers in Stellar
  Interferometry}, volume 5491 of \emph{Society of Photo-Optical
  Instrumentation Engineers (SPIE) Conference Series}, page 842, October 2004.
\newblock \doi{10.1117/12.550581}.

\bibitem[{Labeye} et~al.(2004){Labeye}, {Berger}, {Salhi}, {Demolon},
  {Rousselet-Perraut}, {Malbet}, and {Kern}]{labeye2004}
Pierre~R. {Labeye}, Jean-Philippe {Berger}, Mouna {Salhi}, Pierre {Demolon},
  Karine {Rousselet-Perraut}, Fabien {Malbet}, and Pierre~Y. {Kern}.
\newblock {Integrated optics components in silica on silicon technology for
  stellar interferometry}.
\newblock In Wesley~A. {Traub}, editor, \emph{New Frontiers in Stellar
  Interferometry}, volume 5491 of \emph{Proc. SPIE}, page 667, October 2004.
\newblock \doi{10.1117/12.551863}.

\bibitem[{Minardi} and {Pertsch}(2010)]{minardi2010}
Stefano {Minardi} and Thomas {Pertsch}.
\newblock {Interferometric beam combination with discrete optics}.
\newblock \emph{Optics Letters}, 35\penalty0 (18):\penalty0 3009, August 2010.
\newblock \doi{10.1364/OL.35.003009}.

\bibitem[{Saviauk} et~al.(2013){Saviauk}, {Minardi}, {Dreisow}, {Nolte}, and
  {Pertsch}]{saviauk2013}
Allar {Saviauk}, Stefano {Minardi}, Felix {Dreisow}, Stefan {Nolte}, and Thomas
  {Pertsch}.
\newblock {3D-integrated optics component for astronomical
  spectro-interferometry}.
\newblock \emph{Appl. Opt.}, 52\penalty0 (19):\penalty0 4556, July 2013.
\newblock \doi{10.1364/AO.52.004556}.

\bibitem[{Martin} et~al.(2020){Martin}, {Foin}, {Cassagnettes}, {Ulliac},
  {Courjal}, {Barjot}, {Cvetojevic}, {Vievard}, {Lapeyrere}, {Huby}, and
  {Lacour}]{martin2020}
G.~{Martin}, M.~{Foin}, C.~{Cassagnettes}, G.~{Ulliac}, N.~{Courjal},
  K.~{Barjot}, N.~{Cvetojevic}, S.~{Vievard}, V.~{Lapeyrere}, E.~{Huby}, and
  S.~{Lacour}.
\newblock {Recent results on electro-optic visible multi-telescope beam
  combiner for next generation FIRST/SUBARU instruments: hybrid and passive
  devices}.
\newblock In \emph{Proc. SPIE}, volume 11446 of \emph{Proc. SPIE}, page
  1144626, December 2020.
\newblock \doi{10.1117/12.2562171}.

\bibitem[{Nolte} et~al.(2003){Nolte}, {Will}, {Burghoff}, and
  {Tuennermann}]{Nolte2003}
S.~{Nolte}, M.~{Will}, J.~{Burghoff}, and A.~{Tuennermann}.
\newblock {Femtosecond waveguide writing: a new avenue to three-dimensional
  integrated optics}.
\newblock \emph{Applied Physics A: Materials Science \& Processing},
  77\penalty0 (1):\penalty0 109--111, January 2003.
\newblock \doi{10.1007/s00339-003-2088-6}.

\bibitem[{Gattass} and {Mazur}(2008)]{Gattass2008}
Rafael~R. {Gattass} and Eric {Mazur}.
\newblock {Femtosecond laser micromachining in transparent materials}.
\newblock \emph{Nature Photonics}, 2\penalty0 (4):\penalty0 219--225, April
  2008.
\newblock \doi{10.1038/nphoton.2008.47}.

\bibitem[{Arriola} et~al.(2013){Arriola}, {Gross}, {Jovanovic}, {Charles},
  {Tuthill}, {Olaizola}, {Fuerbach}, and {Withford}]{Arriola2013}
Alexander {Arriola}, Simon {Gross}, Nemanja {Jovanovic}, Ned {Charles},
  Peter~G. {Tuthill}, Santiago~M. {Olaizola}, Alexander {Fuerbach}, and
  Michael~J. {Withford}.
\newblock {Low bend loss waveguides enable compact, efficient 3D photonic
  chips}.
\newblock \emph{Optics Express}, 21\penalty0 (3):\penalty0 2978, February 2013.
\newblock \doi{10.1364/OE.21.002978}.

\bibitem[{Gross} and {Withford}(2015)]{Gross2015}
S.~{Gross} and M.~J. {Withford}.
\newblock {Ultrafast-laser-inscribed 3D integrated photonics: challenges and
  emerging applications}.
\newblock \emph{Nanophotonics}, 4\penalty0 (3):\penalty0 20, November 2015.
\newblock \doi{10.1515/nanoph-2015-0020}.

\bibitem[{Vance} and {Love}(1994)]{vance1994}
R.W.C. {Vance} and J.D. {Love}.
\newblock Design procedures for passive planar coupled waveguide devices.
\newblock \emph{IEE Proceedings - Optoelectronics}, 141:\penalty0 231--241(10),
  August 1994.
\newblock ISSN 1350-2433.
\newblock URL
  \url{https://digital-library.theiet.org/content/journals/10.1049/ip-opt_19941083}.

\bibitem[{Xie} et~al.(2012){Xie}, {Winzer}, {Raybon}, {Gnauck}, {Zhu},
  {Geisler}, and {Edvold}]{xie2012}
Chongjin {Xie}, Peter~J. {Winzer}, Gregory {Raybon}, Alan~H. {Gnauck}, Benyuan
  {Zhu}, Tommy {Geisler}, and Bent {Edvold}.
\newblock {Colorless coherent receiver using 3x3 coupler hybrids and
  single-ended detection}.
\newblock \emph{Optics Express}, 20\penalty0 (2):\penalty0 1164, January 2012.
\newblock \doi{10.1364/OE.20.001164}.

\bibitem[{Hsiao} et~al.(2010){Hsiao}, {Winick}, and {Monnier}]{hsiao2010}
Hsien-Kai {Hsiao}, Kim~A. {Winick}, and John~D. {Monnier}.
\newblock {Midinfrared broadband achromatic astronomical beam combiner for
  nulling interferometry}.
\newblock \emph{Appl. Opt.}, 49\penalty0 (35):\penalty0 6675, December 2010.
\newblock \doi{10.1364/AO.49.006675}.

\bibitem[{Schneider} and {Hattori}(2000)]{schneider2000}
V.~M. {Schneider} and H.~T. {Hattori}.
\newblock {High-tolerance power splitting in symmetric triple-mode evolution
  couplers}.
\newblock \emph{IEEE Journal of Quantum Electronics}, 36\penalty0 (8):\penalty0
  923--930, January 2000.
\newblock \doi{10.1109/3.853545}.

\bibitem[{Le Bouquin} et~al.(2011){Le Bouquin}, {Berger}, {Lazareff}, {Zins},
  {Haguenauer}, {Jocou}, {Kern}, {Millan-Gabet}, {Traub}, {Absil}, {Augereau},
  {Benisty}, {Blind}, {Bonfils}, {Bourget}, {Delboulbe}, {Feautrier},
  {Germain}, {Gitton}, {Gillier}, {Kiekebusch}, {Kluska}, {Knudstrup},
  {Labeye}, {Lizon}, {Monin}, {Magnard}, {Malbet}, {Maurel}, {M{\'e}nard},
  {Micallef}, {Michaud}, {Montagnier}, {Morel}, {Moulin}, {Perraut}, {Popovic},
  {Rabou}, {Rochat}, {Rojas}, {Roussel}, {Roux}, {Stadler}, {Stefl}, {Tatulli},
  and {Ventura}]{pionier2011}
J.~B. {Le Bouquin}, J.~P. {Berger}, B.~{Lazareff}, G.~{Zins}, P.~{Haguenauer},
  L.~{Jocou}, P.~{Kern}, R.~{Millan-Gabet}, W.~{Traub}, O.~{Absil}, J.~C.
  {Augereau}, M.~{Benisty}, N.~{Blind}, X.~{Bonfils}, P.~{Bourget},
  A.~{Delboulbe}, P.~{Feautrier}, M.~{Germain}, P.~{Gitton}, D.~{Gillier},
  M.~{Kiekebusch}, J.~{Kluska}, J.~{Knudstrup}, P.~{Labeye}, J.~L. {Lizon},
  J.~L. {Monin}, Y.~{Magnard}, F.~{Malbet}, D.~{Maurel}, F.~{M{\'e}nard},
  M.~{Micallef}, L.~{Michaud}, G.~{Montagnier}, S.~{Morel}, T.~{Moulin},
  K.~{Perraut}, D.~{Popovic}, P.~{Rabou}, S.~{Rochat}, C.~{Rojas},
  F.~{Roussel}, A.~{Roux}, E.~{Stadler}, S.~{Stefl}, E.~{Tatulli}, and
  N.~{Ventura}.
\newblock {PIONIER: a 4-telescope visitor instrument at VLTI}.
\newblock \emph{Astron. Astrophys}, 535:\penalty0 A67, November 2011.
\newblock \doi{10.1051/0004-6361/201117586}.

\bibitem[{Benisty} et~al.(2009){Benisty}, {Berger}, {Jocou}, {Labeye},
  {Malbet}, {Perraut}, and {Kern}]{benisty2009}
M.~{Benisty}, J.~P. {Berger}, L.~{Jocou}, P.~{Labeye}, F.~{Malbet},
  K.~{Perraut}, and P.~{Kern}.
\newblock {An integrated optics beam combiner for the second generation VLTI
  instruments}.
\newblock \emph{Astron. Astrophys}, 498\penalty0 (2):\penalty0 601--613, May
  2009.
\newblock \doi{10.1051/0004-6361/200811083}.

\bibitem[{Lacour} et~al.(2014){Lacour}, {Tuthill}, {Monnier}, {Kotani},
  {Gauchet}, and {Labeye}]{lacour2014}
S.~{Lacour}, P.~{Tuthill}, J.~D. {Monnier}, T.~{Kotani}, L.~{Gauchet}, and
  P.~{Labeye}.
\newblock {A new interferometer architecture combining nulling with phase
  closure measurements}.
\newblock \emph{Mon. Not. Roy. Astron. Soc.}, 439\penalty0 (4):\penalty0
  4018--4029, April 2014.
\newblock \doi{10.1093/mnras/stu258}.

\bibitem[{Birks}(1992)]{birks1992}
T.~A. {Birks}.
\newblock {Effect of twist in 3 $\times$ 3 fused tapered couplers}.
\newblock \emph{Appl. Opt.}, 31\penalty0 (16):\penalty0 3004--3014, June 1992.
\newblock \doi{10.1364/AO.31.003004}.

\bibitem[Suzuki et~al.(2006)Suzuki, Sharma, Fujimoto, Ippen, and
  Nasu]{Suzuki:06}
Kenya Suzuki, Vikas Sharma, James~G. Fujimoto, Erich~P. Ippen, and Yusuke Nasu.
\newblock Characterization of symmetric 3 $\times$ 3 directional couplers
  fabricated by direct writing with a femtosecond laser oscillator.
\newblock \emph{Opt. Express}, 14\penalty0 (6):\penalty0 2335--2343, Mar 2006.
\newblock \doi{10.1364/OE.14.002335}.
\newblock URL \url{http://www.opticsexpress.org/abstract.cfm?URI=oe-14-6-2335}.

\bibitem[{Spagnolo} et~al.(2013){Spagnolo}, {Vitelli}, {Aparo}, {Mataloni},
  {Sciarrino}, {Crespi}, {Ramponi}, and {Osellame}]{Spagnolo2013}
Nicol{\`o} {Spagnolo}, Chiara {Vitelli}, Lorenzo {Aparo}, Paolo {Mataloni},
  Fabio {Sciarrino}, Andrea {Crespi}, Roberta {Ramponi}, and Roberto
  {Osellame}.
\newblock {Three-photon bosonic coalescence in an integrated tritter}.
\newblock \emph{Nature Communications}, 4:\penalty0 1606, March 2013.
\newblock \doi{10.1038/ncomms2616}.

\bibitem[{Chaboyer} et~al.(2015){Chaboyer}, {Meany}, {Helt}, {Withford}, and
  {Steel}]{Chaboyer2015}
Zachary {Chaboyer}, Thomas {Meany}, L.~G. {Helt}, Michael~J. {Withford}, and
  M.~J. {Steel}.
\newblock {Tunable quantum interference in a 3D integrated circuit}.
\newblock \emph{Scientific Reports}, 5:\penalty0 9601, April 2015.
\newblock \doi{10.1038/srep09601}.

\bibitem[{Lozi} et~al.(2018){Lozi}, {Guyon}, {Jovanovic}, {Takato}, {Singh},
  {Norris}, {Okita}, {Bando}, and {Martinache}]{lozi2018}
Julien {Lozi}, Olivier {Guyon}, Nemanja {Jovanovic}, Naruhisa {Takato}, Garima
  {Singh}, Barnaby {Norris}, Hirofumi {Okita}, Takamasa {Bando}, and Frantz
  {Martinache}.
\newblock {Characterizing vibrations at the Subaru Telescope for the Subaru
  coronagraphic extreme adaptive optics instrument}.
\newblock \emph{Journal of Astronomical Telescopes, Instruments, and Systems},
  4:\penalty0 049001, October 2018.
\newblock \doi{10.1117/1.JATIS.4.4.049001}.

\bibitem[{Vievard} et~al.(2020){Vievard}, {Bos}, {Cassaing}, {Currie}, {Deo},
  {Guyon}, {Jovanovic}, {Keller}, {Lamb}, {Lopez}, {Lozi}, {Martinache},
  {Miller}, {Montmerle-Bonnefois}, {Mugnier}, {N'Diaye}, {Norris}, {Sahoo},
  {Sauvage}, {Skaf}, {Snik}, {Wilby}, and {Wong}]{vievard2020}
S.~{Vievard}, S.~P. {Bos}, F.~{Cassaing}, T.~{Currie}, V.~{Deo}, O.~{Guyon},
  N.~{Jovanovic}, C.~U. {Keller}, M.~{Lamb}, C.~{Lopez}, J.~{Lozi},
  F.~{Martinache}, K.~{Miller}, A.~{Montmerle-Bonnefois}, L.~M. {Mugnier},
  M.~{N'Diaye}, B.~{Norris}, A.~{Sahoo}, J.~F. {Sauvage}, N.~{Skaf}, F.~{Snik},
  M.~J. {Wilby}, and A.~{Wong}.
\newblock {Focal plane wavefront sensing on SUBARU/SCExAO}.
\newblock In \emph{Society of Photo-Optical Instrumentation Engineers (SPIE)
  Conference Series}, volume 11448 of \emph{Society of Photo-Optical
  Instrumentation Engineers (SPIE) Conference Series}, page 114486D, December
  2020.
\newblock \doi{10.1117/12.2562787}.

\bibitem[{Choquet} et~al.(2014){Choquet}, {Menu}, {Perrin}, {Cassaing},
  {Lacour}, and {Eisenhauer}]{choquet2014}
{\'E}.~{Choquet}, J.~{Menu}, G.~{Perrin}, F.~{Cassaing}, S.~{Lacour}, and
  F.~{Eisenhauer}.
\newblock {Comparison of fringe-tracking algorithms for single-mode
  near-infrared long-baseline interferometers}.
\newblock \emph{Astron. Astrophys}, 569:\penalty0 A2, September 2014.
\newblock \doi{10.1051/0004-6361/201220223}.

\bibitem[{Travouillon} et~al.(2009){Travouillon}, {Els}, {Riddle},
  {Sch{\"o}ck}, and {Skidmore}]{travouillon2009}
Tony {Travouillon}, Sebastian {Els}, Reed~L. {Riddle}, Matthias {Sch{\"o}ck},
  and Warren {Skidmore}.
\newblock {Thirty Meter Telescope Site Testing VII: Turbulence Coherence Time}.
\newblock \emph{Publications of the Astronomical Society of the Pacific},
  121\penalty0 (881):\penalty0 787, July 2009.
\newblock \doi{10.1086/605295}.

\bibitem[{Coud{\'e} du Foresto} et~al.(2000){Coud{\'e} du Foresto},
  {Faucherre}, {Hubin}, and {Gitton}]{coude2000}
V.~{Coud{\'e} du Foresto}, M.~{Faucherre}, N.~{Hubin}, and P.~{Gitton}.
\newblock {Using single-mode fibers to monitor fast Strehl ratio fluctuations.
  Application to a 3.6 m telescope corrected by adaptive optics}.
\newblock \emph{Astron. Astrophyss}, 145:\penalty0 305--310, August 2000.
\newblock \doi{10.1051/aas:2000351}.

\bibitem[{Currie} et~al.(2020){Currie}, {Guyon}, {Lozi}, {Sahoo}, {Vievard},
  {Deo}, {Chilcote}, {Groff}, {Brandt}, {Lawson}, {Skaf}, {Martinache}, and
  {Kasdin}]{currie2020}
Thayne {Currie}, Olivier {Guyon}, Julien {Lozi}, Ananya {Sahoo}, Sebastien
  {Vievard}, Vincent {Deo}, Jeffrey {Chilcote}, Tyler {Groff}, Timothy~D.
  {Brandt}, Kellen {Lawson}, Nour {Skaf}, Frantz {Martinache}, and N.~Jeremy
  {Kasdin}.
\newblock {On-sky performance and recent results from the Subaru coronagraphic
  extreme adaptive optics system}.
\newblock In \emph{Society of Photo-Optical Instrumentation Engineers (SPIE)
  Conference Series}, volume 11448 of \emph{Society of Photo-Optical
  Instrumentation Engineers (SPIE) Conference Series}, page 114487H, December
  2020.
\newblock \doi{10.1117/12.2576349}.

\bibitem[{Mennesson} et~al.(2011){Mennesson}, {Hanot}, {Serabyn}, {Liewer},
  {Martin}, and {Mawet}]{Mennesson2011}
B.~{Mennesson}, C.~{Hanot}, E.~{Serabyn}, K.~{Liewer}, S.~R. {Martin}, and
  D.~{Mawet}.
\newblock {High-contrast Stellar Observations within the Diffraction Limit at
  the Palomar Hale Telescope}.
\newblock \emph{Astrophys. J}, 743\penalty0 (2):\penalty0 178, December 2011.
\newblock \doi{10.1088/0004-637X/743/2/178}.

\bibitem[{K{\"u}hn} et~al.(2015){K{\"u}hn}, {Mennesson}, {Liewer}, {Martin},
  {Loya}, {Millan-gabet}, and {Serabyn}]{kuhn2015}
J.~{K{\"u}hn}, B.~{Mennesson}, K.~{Liewer}, S.~{Martin}, F.~{Loya},
  R.~{Millan-gabet}, and E.~{Serabyn}.
\newblock {Exploring Intermediate (5-40 au) Scales around AB Aurigae with the
  Palomar Fiber Nuller}.
\newblock \emph{Astrophys. J}, 800\penalty0 (1):\penalty0 55, February 2015.
\newblock \doi{10.1088/0004-637X/800/1/55}.

\bibitem[{Serabyn} et~al.(2019){Serabyn}, {Mennesson}, {Martin}, {Liewer}, and
  {K{\"u}hn}]{serabyn2019}
E.~{Serabyn}, B.~{Mennesson}, S.~{Martin}, K.~{Liewer}, and J.~{K{\"u}hn}.
\newblock {Nulling at short wavelengths: theoretical performance constraints
  and a demonstration of faint companion detection inside the diffraction limit
  with a rotating-baseline interferometer}.
\newblock \emph{Mon. Not. Roy. Astron. Soc.}, 489\penalty0 (1):\penalty0
  1291--1303, October 2019.
\newblock \doi{10.1093/mnras/stz2163}.

\end{thebibliography}

\end{document}